\documentclass[runningheads]{llncs}

\usepackage[T1]{fontenc}
\usepackage{cite}
\usepackage{amsmath,amssymb,amsfonts}
\usepackage{graphicx}
\usepackage{textcomp}
\usepackage{xcolor}
\usepackage{listings}
\usepackage{float}
\usepackage{lipsum}
\usepackage{lineno} 
\usepackage{algpseudocode} 
\usepackage[ruled,vlined]{algorithm2e}
\usepackage{epsfig}
\usepackage[export]{adjustbox}
\usepackage{multirow}
\usepackage{graphicx}
\usepackage{subfig}
\usepackage{float}
\usepackage{subcaption}

\def\BibTeX{{\rm B\kern-.05em{\sc i\kern-.025em b}\kern-.08em
    T\kern-.1667em\lower.7ex\hbox{E}\kern-.125emX}}
\begin{document}

\titlerunning{ML Inference: Parallel Batch Processing with Serverless Functions}

\title{Scalable and Cost-Efficient ML Inference: Parallel Batch Processing with Serverless Functions}

\author{
    Amine Barrak \and
    Emna Ksontini
}
\authorrunning{A. Barrak and E. Ksontini}

\institute{
    Department of Computer Science and Engineering \\Oakland University, Michigan, USA\\
    \email{\{aminebarrak, emnaksontini\}@oakland.edu}
}

\newcommand{\al}{\textit{et al.}}
\newcommand{\cf}{{\textit{cf.,}}}
\newcommand{\aka}{{\textit{a.k.a.,}}}
\newcommand{\smallsection}[1]{\noindent\textbf{#1.}}
\newcommand{\eg}{{\textit{e.g.,}}}
\newcommand{\ie}{{\textit{i.e.,}}}

\maketitle

\begin{abstract}
As data-intensive applications grow, batch processing in limited-resource environments faces scalability and resource management challenges. Serverless computing offers a flexible alternative, enabling dynamic resource allocation and automatic scaling. This paper explores how serverless architectures can make large-scale ML inference tasks faster and cost-effective by decomposing monolithic processes into parallel functions. Through a case study on sentiment analysis using the DistilBERT model and the IMDb dataset, we demonstrate that serverless parallel processing can reduce execution time by over 95\% compared to monolithic approaches, at the same cost.


\end{abstract}

\section{Introduction}
\label{sec:introduction} 

The rapid growth of data-intensive applications, such as machine learning and real-time processing, demands substantial computational resources, driving innovations in processing and infrastructure \cite{BONNER2017253}. In resource-limited environments, traditional batch processing frameworks often become bottlenecks due to fixed resource allocation, leading to inefficiencies like underutilization during low workloads or resource exhaustion during peak demand \cite{henning2024benchmarking}. Sequential batch processing in these setups further constrains scalability, and scaling typically requires costly infrastructure investments \cite{deng2024cloud}.

In this context, serverless computing has emerged as a flexible and scalable alternative \cite{adzic2017serverless}. By decoupling resource management from application logic, serverless architectures enable automatic scaling based on workload, reducing overhead and potentially lowering costs \cite{liu2023demystifying}. However, the benefits of serverless batch processing depend heavily on effective architecture design \cite{worah2024serverless}. Transitioning from monolithic to serverless isn’t a simple lift-and-shift; it requires rethinking task decomposition and execution. A common mistake is moving a monolithic batch task to a serverless setup without restructuring \cite{mcgrath2017serverless}. Serverless platforms excel at parallel execution, but without breaking tasks into smaller units, the benefits of parallelism are lost \cite{mampage2022holistic}, leading to inefficient sequential execution, scalability issues, and higher latency \cite{werner2024reference}. Additionally, the economics of serverless computing introduce a performance-cost trade-off. While dividing tasks into smaller batches can reduce execution time, triggering numerous functions simultaneously can increase costs due to the pay-per-invocation pricing model \cite{barrak}. Thus, balancing performance and cost efficiency requires careful consideration.

This paper explores how serverless architectures can enhance ML inference by transforming monolithic batch-processing tasks into smaller, parallel functions optimized for serverless execution. Inference involves applying a trained model to new data using forward propagation. Although batch processing improves efficiency through hardware parallelism, resource-limited environments and poor task decomposition often create bottlenecks. We demonstrate that breaking tasks into parallel functions can reduce latency and optimize resource use in serverless environments\footnote{\url{https://tinyurl.com/ICSOC2024}}. This paper makes the following key contributions: 
\begin{itemize} 
\item We demonstrate how to transform monolithic ML inference tasks into parallel functions optimized for serverless execution. 
\item We analyze trade-offs between performance and cost in serverless batch processing, offering insights on when serverless computing is faster and more cost-effective for large-scale ML tasks. 
\end{itemize}

\section{Related Work}

Serverless computing has emerged as a powerful paradigm for data-intensive applications, enabling scalable and cost-efficient processing through parallel execution. Giménez Alventosa et al. \cite{gimenez2019framework} highlight AWS Lambda’s effectiveness in executing MapReduce jobs, demonstrating that serverless architectures are well-suited for loosely coupled, parallelizable workloads. Similarly, Aytekin and Johansson \cite{aytekin2019exploiting} observed significant performance gains in large-scale optimization problems using AWS Lambda, underscoring the potential speed advantages of parallel processing within serverless environments.

Task decomposition and parallelization strategies are critical for enhancing the scalability and performance of data-intensive applications. Carreira et al. \cite{carreira2018serverless} and Jiang et al. \cite{jiang2021towards} demonstrated that serverless platforms can improve performance in machine learning tasks when these tasks are broken down into smaller, parallel units, reducing overhead and optimizing resource use. Barrak et al. \cite{barrak2023exploring} further addressed cost-performance trade-offs in serverless training environments, introducing the SPIRT architecture—a peer-to-peer serverless approach designed to distribute training loads \cite{barrak2023spirt}. Recognizing that decomposing functions can introduce communication overhead, Barrak proposed in-database computations \cite{barrak2024incorporating} to reduce costs and enhance scalability.

While existing studies provide general solutions for parallelization and task decomposition, they often overlook the unique demands of batch-processing in machine learning. Our work builds on these advances by offering a tailored approach to decomposing monolithic ML inference tasks for optimized serverless execution, analyzing the trade-offs in performance and cost specifically for batch-processing in serverless environments.

\section{System Design and Methodology}
\label{sec:study}

Batch processing improves use and efficiency with large datasets, but sequential execution in limited environments creates scalability challenges.

\subsection{Monolithic vs. Parallel Function Execution} 
\begin{figure}[H]
 \centering
 \includegraphics[scale=0.5]{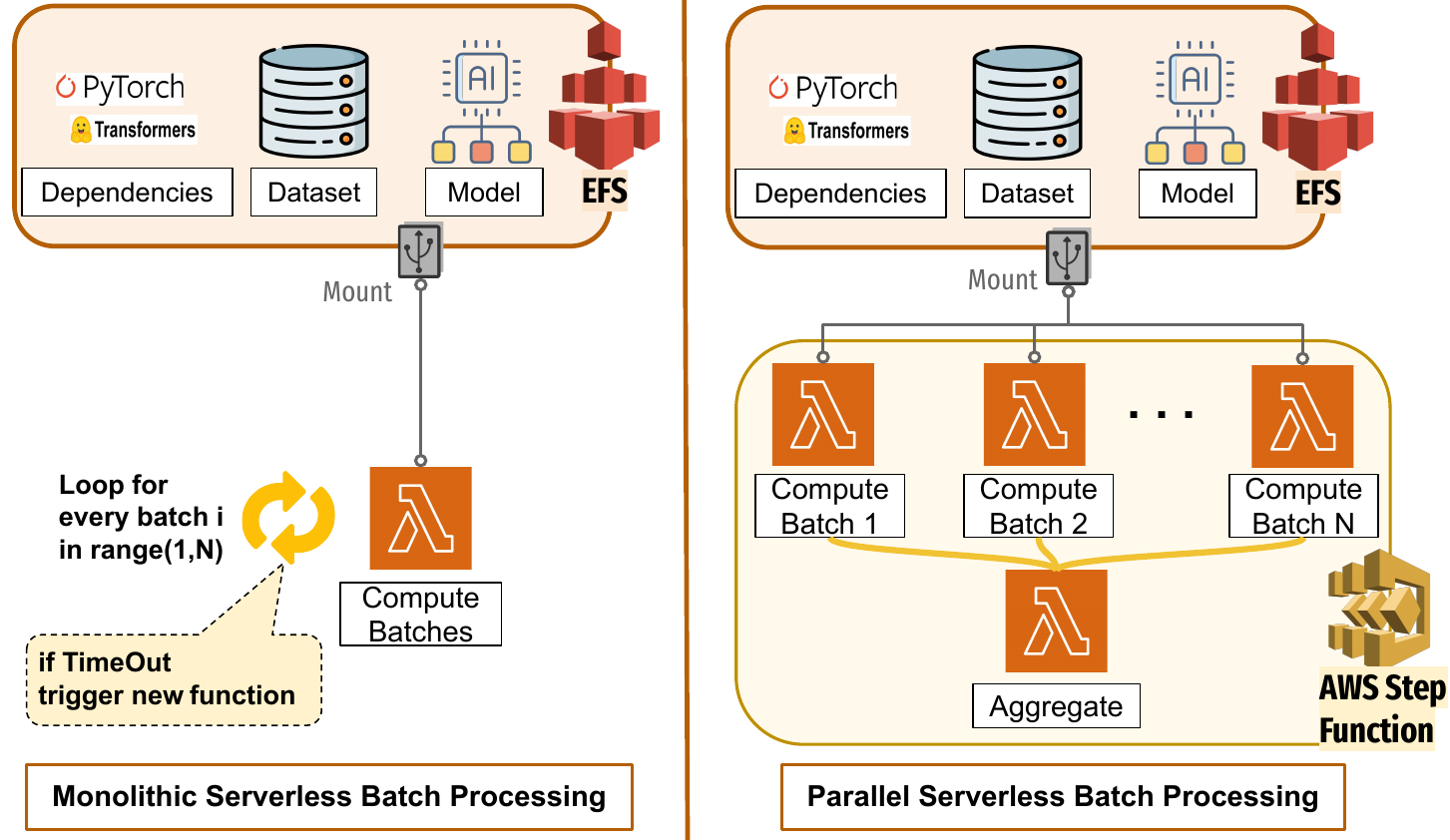}
 \caption{Comparison of Monolithic and Parallel-Function Batch Processing Workflows in Serverless Computing}
 \label{fig:arch}
 \vspace{-10pt}
\end{figure}

To understand the impact of batch size on serverless processing, we implemented two approaches: monolithic and parallel batch processing, as shown in Figure \ref{fig:arch}.

\noindent\textbf{Monolithic-Function Batch Processing}
In monolithic processing, a single serverless function handles the entire batch. After each epoch, the function checks whether enough time remains to process another batch or if a new function should be triggered. This cycle continues until all batches are processed.\\

\noindent\textbf{Parallel Function Batch Processing}
 In parallel processing, the batch is split into smaller chunks, each processed by a separate serverless function. AWS Step Functions orchestrate these tasks, running up to 10 functions concurrently by default. As one function completes, another is triggered. AWS also allows requesting an increase in concurrency limits as needed.

To address the limitations of serverless functions, such as their short-lived nature and dependency management challenges, we used Amazon Elastic File System (EFS). EFS stored the model, dataset, and dependencies, allowing each function to access them during execution, enabling more complex, resource-intensive tasks to be processed efficiently.

\subsection{Evaluation Metrics }

\textbf{Execution Time (min)}
 The total time required to process the batch, measured from the start of the first function invocation to the completion of the final result.
 
\noindent\textbf{Maximum RAM Used (MB)}
The maximum amount of memory utilized by the serverless function(s) during the processing of the batch. This value is recorded and reported by the function during its execution.

\noindent\textbf{Computation Cost in (\$)}
The cost for running serverless functions is computed based on execution time, RAM used, and the number of invocations. The following equations calculate the cost for both monolithic and parallel function approaches:
\begin{small}
\begin{align}
    \text{Cost}_{\text{parallel}} &= \sum_{i=1}^{n} \text{Execution Time}_i \times \text{Price}_i + \text{Step Functions Cost} \\
    \text{Cost}_{\text{monolithic}} &= \text{Execution Time (ms)} \times \text{Price per 1ms at RAM Used}
\end{align}
\end{small}
        
Here, \( n \) is the total number of parallel functions invoked.

\section{Case Study: Sentimental Analysis}
\label{sec:results}

To evaluate our approach, We conducted a case study on sentiment analysis using a serverless model to evaluate the effectiveness of batch processing in serverless environments, particularly for large-scale NLP tasks. The specifics are as follows:
\begin{itemize} \item \textbf{Model:} DistilBERT, a distilled version of BERT with 66 million parameters, was used for sentiment analysis, classifying text as positive or negative. \item \textbf{Dataset:} The IMDb dataset, containing 25,000 balanced movie reviews, was used for sentiment classification. 
\end{itemize}
\vspace{-5mm}
\subsection{Methodology and Experimental Setup} We conducted experiments focused on ML inference, varying batch sizes (50, 100, 125, 200, 250, 333, 500, 625, and 1000), where larger batch sizes required fewer total batches to process the dataset. We employed two processing methods:
 \begin{itemize} \item \textit{Monolithic Processing:} Batches were processed sequentially using a single serverless function, simulating a traditional step-by-step flow. \item \textit{Parallel Processing:} Multiple serverless functions processed batches simultaneously, leveraging serverless scalability to reduce execution time. \end{itemize}

\begin{figure*}[t!]
    \centering
    \begin{minipage}[t]{0.5\textwidth}
        \centering
        \includegraphics[width=\textwidth]{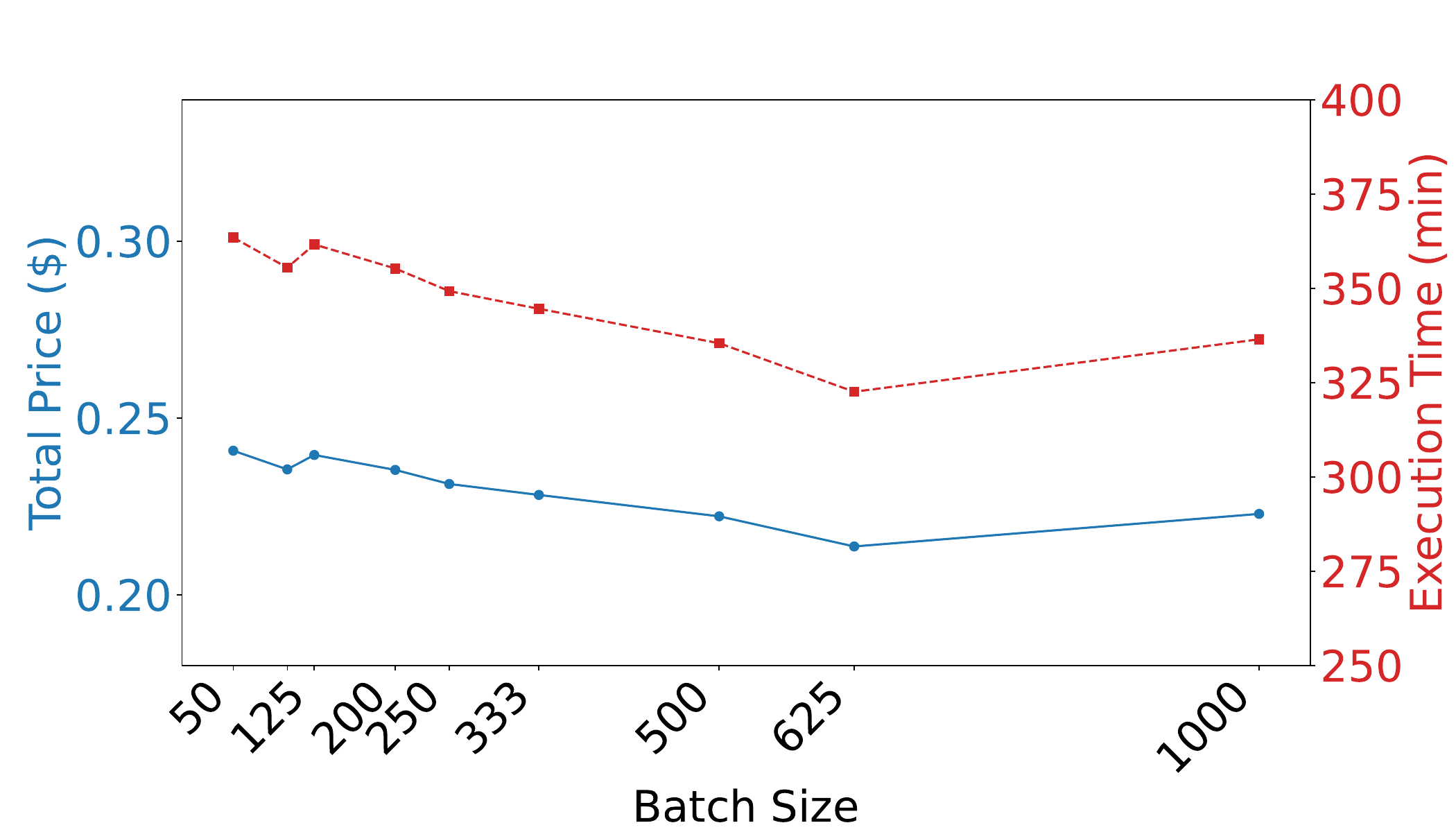}
        \caption*{(a) Monolithic Processing}
        \label{fig:sequential_batch}
    \end{minipage}%
    \hfill
    \begin{minipage}[t]{0.5\textwidth}
        \centering
        \includegraphics[width=\textwidth]{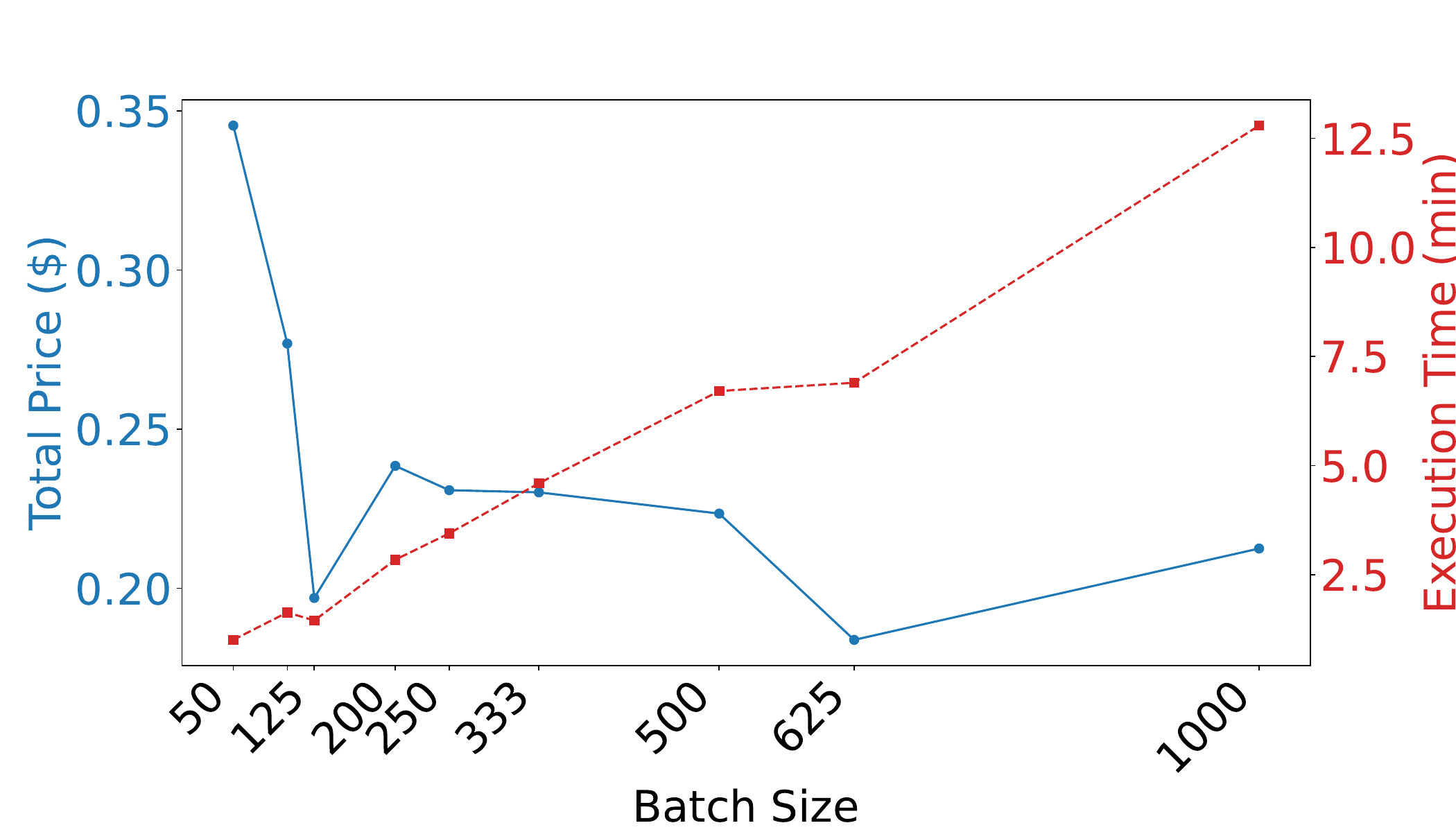}
        \caption*{(b) Parallel Processing}
        \label{fig:parallel_batch}
    \end{minipage}
    \caption{Price and Execution Time for Monolithic vs. Parallel Batch Processing.}
    \label{fig:batch_processing_comparison}
\end{figure*}

For each batch size and method, we measured execution time and cost. The results are presented in Figure \ref{fig:batch_processing_comparison}.

\vspace{-3mm}
\subsection{Experimental Results}

Figure \ref{fig:batch_processing_comparison}(a) shows the analysis of monolithic batch processing, where both cost and execution time remain relatively stable across batch sizes. The total cost slightly decreases from \$0.2408 at a batch size of 50 to \$0.2229 at a batch size of 1000. Similarly, execution time decreases modestly from 363.5 minutes (6.06 hours) to 336.5 minutes (5.61 hours). This is due to larger batch sizes requiring fewer function invocations, reducing overhead.

In contrast, Figure \ref{fig:batch_processing_comparison}(b) illustrates the dynamic nature of parallel batch processing. For smaller batches, execution time significantly decreases, reaching 1.01 minutes for a batch size of 50, enabled by 500 parallel functions. However, this speed comes with a peak cost of \$0.3454 for the smallest batch size. As batch sizes increase, cost stabilizes around \$0.1838 for batches of 500 to 625, while execution time increases slightly but remains under 12.79 minutes.

A direct comparison shows the significant advantages of parallel processing, especially for ML inference tasks where speed is critical. For a similar cost range (\$0.20 to \$0.25), monolithic processing takes 325 to 375 minutes (5.42 to 6.25 hours), while parallel processing completes the same tasks in just 2.5 to 7.5 minutes—a reduction of over 95\%.

The cost-effectiveness of parallel processing is due to how serverless platforms charge for compute time. Both approaches used the same amount of RAM (830-850 MB), as inference tasks load model parameters into memory without requiring additional space for training-specific processes like back-propagation. This stability in memory usage ensures consistent costs across both methods, but parallel processing reduces execution time by leveraging concurrency.

\section{Conclusion}
\label{sec:conclusion}
In this paper, we have explored the potential and challenges of leveraging serverless computing for batch processing in machine learning (ML) inference tasks. By comparing monolithic and parallel-function execution, we demonstrated that monolithic processing is inefficient in scaling for large datasets, resulting in longer execution times. In contrast, parallel-function processing, orchestrated within serverless environments, significantly reduces execution time while maintaining similar cost. Our findings reveal that migrating monolithic tasks to serverless without restructuring preserves inefficiencies, highlighting the importance of breaking down tasks to take full advantage of serverless benefits.

\bibliographystyle{IEEEtran}

\bibliography{bibliography}
\end{document}